\documentclass[twocolumn,pra,a4paper,nofootinbib,showpacs,aps,floatfix,superscriptaddress]{revtex4-1}
\usepackage[utf8]{inputenc}
\usepackage{graphicx}
\usepackage{bm}
\usepackage{amsmath}
\usepackage{color}

\begin{document}

\title{Invariant-based inverse engineering for fast and robust load transport in a double pendulum bridge crane}
\author{I. Lizuain}
\email{ion.lizuain@ehu.eus}
\affiliation{Department of Applied Mathematics, University of the Basque Country UPV/EHU, Donostia-San Sebastian, Spain}
\author{A. Tobalina}
\affiliation{Department of Physical Chemistry, University of the Basque Country UPV/EHU, Apdo. 644, Bilbao, Spain}
\author{A. Rodriguez-Prieto}
\affiliation{Departament of Applied Mathematics, University of the Basque Country UPV/EHU, Bilbao, Spain}
\author{J. G. Muga}
\affiliation{Department of Physical Chemistry, University of the Basque Country UPV/EHU, Apdo. 644, Bilbao, Spain}

\begin{abstract}
We set a shortcut-to-adiabaticity strategy to design the trolley motion in a double-pendulum bridge crane. 
The trajectories found guarantee payload transport without residual excitation regardless of the initial conditions 
within the small oscillations regime. The results are compared with exact dynamics to set the working domain of the approach. 
The method is free from instabilities due boundary effects  or to resonances with the two natural frequencies.    
\end{abstract}

\maketitle

%
%
%
\section{Introduction}
The concept of adiabaticity is ubiquitous in physics, but it is not fully exploited in 
mechanical engineering and control applications.  
Adiabatic theorems set  the existence of approximate adiabatic invariants, such as the action integral in classical mechanics, 
when the control parameters of a given physical system
vary slowly enough in time  \cite{Sakurai1993}. 

Adiabaticity is often used to drive systems  in a robust manner. 
An example is a load hanging as a simple pendulum from a moving trolley on a bridge crane.   
If the trolley travels slowly enough between two points, the energy of the pendulum is an adiabatic invariant and stays constant for different smooth trolley trajectories for  the same initial and final points.   
In particular, the minimum energy configuration, in which the oscillating mass stays at relative rest with respect to the suspension point,
is preserved.  More generally, for other initial states the final energy will not suffer excitations.  
However, the intrinsic slowness of such  processes may be problematic, either because  long operation times are impractical, or because during a long process time the ideal dynamics can be affected by the accumulation of 
random and/or uncontrollable perturbations that spoil the desired result.  

To overcome these problems, ``Shortcuts To Adiabaticity'' (STA) methods have been developed in the last decade.
The idea is to reach the same results of an adiabatic protocol in short times  \cite{review2013,review2019}. 
In STA, the adiabatic invariant is not kept constant throughout the process, but the initial value is recovered at final time. For the 
simple example of the load hanging from a moving trolley, the shortcuts are certain special and fast driving trajectories  
of the trolley that induce transitory excitations, but leave the load at final time with the same energy it had initially.

\begin{figure}[t]
\begin{center}
\includegraphics[width=7cm]{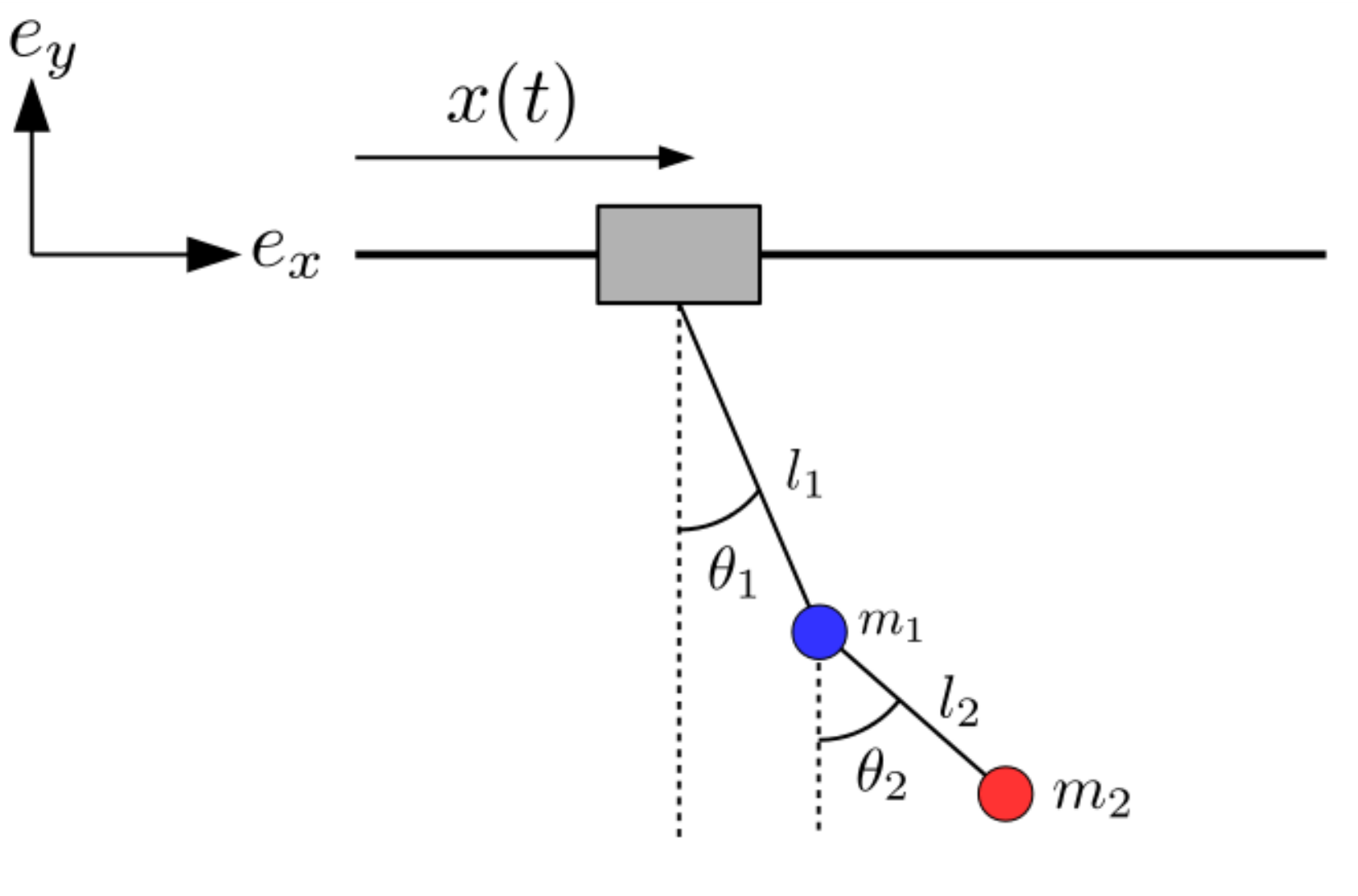}
\caption{Double pendulum overhead crane scheme and relevant physical parameters.}
\label{fig_crane}
\end{center}
\end{figure}

STA methods have been succesfully applied to many different fields and processes
in quantum systems,    
such as: 
quantum computation \cite{sarandy2011,palmero2017,delcampo2012,takahashi2017},
cooling \cite{onofrio2017},
quantum transport \cite{torrontegui2011,bowler2012},
quantum state preparation \cite{chen2010b,bason2012,zhang2013,zhou2017}, 
manipulation of cold atoms \cite{torrontegui2012,rohringer2015,schaff2010,schaff2011,torrontegui2013,kiely2013}, 
control of polyatomic molecules \cite{mashuda2015}.
They have been also applied to design optical devices \cite{chen2012,garaot2014},
and recently in mechanical engineering  to design fast and robust protocols 
to control overhead cranes \cite{Torrontegui2017,Gonzalez-Resines2017}. 
Perhaps surprisingly, because of the differing orders of magnitude involved, the physics of crane control are much related, in some formal aspects and domains even identical,  
to the physics of microscopic particle transport in moving traps 
\cite{torrontegui2011}.
In both domains the linearized models imply a moving harmonic oscillator. When the setting is more realistic though, beyond the simplest scenarios, 
the models become specific and require specialized treatment as  in the system addressed  in this paper, see  Fig. \ref{fig_crane}, a planar, double-pendulum,  
hook ($m_1$) and load ($m_2$) system suspended from  a moving trolley.      
This is a relevant model as cranes behave like moving double pendulums due to different reasons, for example the large scale of the payload, 
or weighty hooks \cite{Masoud2014,Sun2017}. The control of a moving double pendulum dynamics is significantly more difficult than the single pendulum, 
with  two  unactuated degrees of freedom (angles $\theta_1$ and $\theta_2$ in 
Fig. \ref{fig_crane}) and only one actuator (the trolley position $x$). 
Compared to studies on single pendulum cranes, this system is much less explored, for a recent brief review on recent papers and approaches applied see \cite{Sun2017}. 
Control approaches with and without feedback have been worked out 
and their pros and cons have been well discussed \cite{Masoud2014,Sun2017,Vaughan2010}. 
Our STA approach is presented here in an elementary way without feedback  
but it may be adapted and incorporated into methods with feedback as well.      

Among the different STA approaches, dynamical invariant based inverse engineering is one of the most successful
and is the one followed here. 
The essence of the method is to identify exact (rather than adiabatic) dynamical invariants, 
set boundary conditions to cancel final excitation, design the dynamics compatible with these conditions, and deduce the necessary controls  from the dynamics  
thanks to the relations between dynamical invariants and  Hamiltonian.   
For the moving double pendulum the STA consists on designing trolley trajectories $x(t)$ from $x=0$ to  $x=d$ so that the system ends up at final process time $t_f$ 
without excitations. From the point of view of STA process design, this system poses interesting, non-trivial challenges with respect to the single pendulum, as we shall see.

The article is organized as follows.
The physical model and Hamiltonian of the system are set in Sec. \ref{sec_model}, both in exact form and in the small oscillation regime. 
Dynamically decoupled normal modes are found in Sec. \ref{sec_normalmodes}, and then
the STA protocol is designed in Sec. \ref{sec_sta}. 
Numerical results are presented in Sec. \ref{sec_results} and, 
finally, in Sec. \ref{sec_conclussions} we end with the  conclusions and discuss some open questions.

\section{Physical model}
\label{sec_model}
The physical model and relevant parameters are shown in Fig.  \ref{fig_crane}.
The model assumes several conditions and idealizations: 
{ (i)} the mass of the wires and friction are neglected;
{ (ii)} point masses; 
{ (iii)} constant wire lengths $l_1$ and $l_2$; 
{ (iv)} the trolley position is treated as a control parameter rather than a dynamical
variable. This last assumption is a common and simplifying assumption \cite{Abdel-Rahman2003} that requires a good controller,  
but a more fundamental approach considering  the trolley as a dynamical variable is also possible as in \cite{Torrontegui2017}.

\subsection{Lagrangian}
In terms of the angles $\theta_1$ and $\theta_2$, see Fig.  \ref{fig_crane}, the Cartesian coordinates of each mass 
in a rest frame are given by 
 \begin{eqnarray}
 x_1=x+l_1\sin\theta_1,&\quad&  y_1=-l_1\cos\theta_1,
\nonumber\\
 x_2=x_1+l_2\sin\theta_2,&\quad& y_2=y_1-l_2\cos\theta_2,
\end{eqnarray}
so kinetic $(T)$ and potential $(V)$  energies are given by 
\begin{eqnarray}
 T&=&\frac{1}{2}m_1(\dot x_1^2+\dot y_1^2)+\frac{1}{2}(\dot x_2^2+\dot y_2^2),
 \nonumber\\
 V&=&m_1gy_1+m_2gy_2,
 \label{exact_TV}
\end{eqnarray}
where the dots represent time derivatives. The Lagrangian of the system using $\theta_1$ and $\theta_2$ as 
generalized coordinates and dynamical variables 
and $x(t)$ as a control parameter will be given by 
\begin{eqnarray}
 L&=&L(\theta_i,\dot\theta_i;t)=T-V.
 \label{lagrangian_exact}
\end{eqnarray}
%
%
To avoid deformations or excessive tensions, cranes usually work in the small oscillations regime,
in which  $\theta_i$ are small so that we may approximate
$\sin\theta_i\approx \theta_i$ and $\cos\theta_i\approx 1-{\theta_i^2}/{2}$. Angular velocities $\dot\theta$ will be considered small as well.
This approximation linearizes the dynamical equations of motion of the system. 
Results found  with exact and approximate dynamics will be compared  later to check the validity of the approximation and its limits.

In this small oscillation regime and keeping up to second order quadratic terms in $\theta_i$ and $\dot\theta_i$, 
kinetic and potential energies are given by
\begin{eqnarray}
 T&\approx&\frac{1}{2}M \dot x^2+
 \frac{1}{2}(\dot\theta_1,\dot\theta_2)
 \left(\begin{array}{cc}
 M l_1^2 &  m_2l_1 l_2 \\
  m_2l_1 l_2 &  m_2 l_2^2 
\end{array}\right)
\left(\begin{matrix}\dot \theta_1\\ \dot \theta_2\end{matrix}\right)
\nonumber\\
&+& \dot x\left(M l_1, m_2l_2\right)
 \left(\begin{matrix} \dot \theta_1\\ \dot \theta_2\end{matrix}\right),
 \nonumber\\
V&\approx&- M g l_1\left(1-\frac{\theta_1^2}{2}\right)-m_2 g l_2 \left(1-\frac{\theta_2^2}{2}\right),
\end{eqnarray}
where $M$ denotes the total mass $M=m_1+m_2$. 
and the Lagrangian becomes
\begin{eqnarray}
 L&\approx&T-V\nonumber\\
 &=&\frac{1}{2}(\dot\theta_1,\dot\theta_2)
 \left(\begin{array}{cc}
 M l_1^2 &  m_2l_1 l_2 
 \\
  m_2l_1 l_2 &  m_2 l_2^2 
\end{array}\right)
 \left(\begin{matrix}\dot \theta_1\\ \dot \theta_2\end{matrix}\right)
 \nonumber\\
 &-& \frac{1}{2}(\theta_1,\theta_2)
 \left(\begin{array}{cc}
 Mgl_1 & 0 \\
 0 &  m_2g l_2
\end{array}\right)
 \left(\begin{matrix} \theta_1\\ \theta_2\end{matrix}\right)
 \nonumber\\
  &+& \dot x\left(M l_1, m_2l_2\right)
 \left(\begin{matrix} \dot \theta_1\\ \dot \theta_2\end{matrix}\right),
  \label{lagrangian_approx}
\end{eqnarray}
where purely time-dependent and constant terms have been omitted since they do not affect the dynamics.
%
%
%
\subsection{Hamiltonian}
To implement a Hamiltonian formulation, which is more convenient to treat the invariants
and inverse engineering of trolley trajectories, 
we need the conjugate momentum of each $\theta_i$, 
\begin{eqnarray}
 p_{\theta_1}&=&\frac{\partial L_\theta}{\partial \dot \theta_1}=M l_1\dot x+M l_1^2\dot\theta_1+m_2l_1l_2\dot \theta_2,
\nonumber \\
 p_{\theta_2}&=&\frac{\partial L_\theta}{\partial \dot \theta_2}=m_2l_2\dot x+ m_2l_1 l_2  \dot\theta_1+ m_2 l_2^2\dot \theta_2.
\end{eqnarray}
These relations can be inverted to have the generalized velocities $\dot\theta_i$ in terms of the generalized momenta $p_{\theta_i}$. 
The Hamiltonian is found from the Lagrangian as 
\begin{eqnarray}
H_\theta&=&\sum_{i=1}^2 \dot\theta_ip_{\theta_i}-L
\nonumber\\
&=&\frac{p_{\theta_1}^2}{2m_1l_1^2}+\frac{p_{\theta_2}^2}{2\mu l_2^2}-\frac{p_{\theta_1}p_{\theta_2}}{m_1l_1l_2}
\nonumber\\
 &+&\frac{1}{2}Mgl_1\theta_1^2+\frac{1}{2}m_2gl_2\theta_2^2-\dot x \frac{p_{\theta_1}}{l_1},
\end{eqnarray}
where $\mu=\frac{m_1m_2}{m_1+m_2}$ is the reduced mass and where  
constant terms that do not affect the dynamics have been neglected.

In matrix representation the Hamiltonian can be written as
\begin{eqnarray}
 H_\theta&=&
 \frac{1}{2}(p_{\theta_1},p_{\theta_2})T \left(\begin{matrix}p_{\theta_1}\\ p_{\theta_2}\end{matrix}\right)
 + \frac{1}{2}(\theta_1,\theta_2) K \left(\begin{matrix} \theta_1\\ \theta_2\end{matrix}\right)\nonumber\\
  &-&\left(\frac{\dot x }{l_1},0\right)\left(\begin{matrix} p_{\theta_1}\\ p_{\theta_2}\end{matrix}\right),
\end{eqnarray}
where 
%
$$
\hspace*{-.1cm} 
T=\left(\begin{array}{cc} \frac{1}{m_1 l_1^2 } & \frac{-1}{m_1 l_1 l_2 } 
\\ \frac{-1}{m_1 l_1 l_2 } &  \frac{m_1+m_2}{m_1 m_2  l_2^2 }\end{array}\right)\!
\textrm{; }
 K=\left(\begin{array}{cc} Mgl_1 & 0 \\ 0 &  m_2g l_2\end{array}\right).
$$
%
Whereas the potential matrix $K$ is diagonal, the kinetic matrix $T$ is not, i. e., $p_{\theta_1}$ and $p_{\theta_2}$ momenta are coupled.
We want to find a coordinate transformation, i. e., normal modes,  where both the coordinates and momenta are uncoupled so that 
we can easily get  the dynamical invariants
to inverse engineer $x(t)$. In the following section, these normal modes will be calculated
following \cite{lizuain2017}. Normal modes for the double pendulum with fixed suspension point are known \cite{Blevins1979}, but 
our treatment takes the motion of the trolley into account. 
Finding dynamical normal modes for quadratic {\it time-dependent} Hamiltonians is generically non-trivial \cite{lizuain2017}, but in 
this system the task is facilitated by the fact that the time-dependence appears in linear terms via $\dot{x}(t)$.

\section{Normal modes}
\label{sec_normalmodes}
\subsection{Diagonalization of $H_\theta$}
%
Let us first define a new set of coordinates $u_1$ and $u_2$ by the linear transformation
\begin{equation}
 \left(\begin{matrix} u_1\\ u_2\end{matrix}\right)=
 A \left(\begin{matrix} \theta_1\\ \theta_2\end{matrix}\right),
\end{equation}
where the $A$ matrix is yet to be determined. The corresponding momenta transform according to  \cite{lizuain2017}
\begin{equation}
 \left(\begin{matrix} p_{u_1}\\ p_{u_2}\end{matrix}\right)=
 A^{-T} \left(\begin{matrix} p_{\theta_1}\\ p_{\theta_2}\end{matrix}\right),
\end{equation}
where $A^{-T}=(A^{-1})^T$ stands for the transpose of the inverse matrix. The Hamiltonian in these variables reads
\begin{eqnarray}
 H_u&=&
 \frac{1}{2}(p_{u_1},p_{u_2})\left(A T A^T\right) \left(\begin{matrix}p_{u_1}\\ p_{u_2}\end{matrix}\right)\nonumber\\
 &+& \frac{1}{2}(u_1,u_2) \left(A^{-T}K A^{-1}\right) \left(\begin{matrix} u_1\\ u_2\end{matrix}\right)\nonumber\\
 &-&\left(\frac{\dot x }{l_1},0\right)A^T\left(\begin{matrix} p_{u_1}\\ p_{u_2}\end{matrix}\right).
 \label{Hu_coupled}
\end{eqnarray}
We now look for a transformation matrix $A$ that diagonalizes simultaneously both the $A T A^T$ and  $A^{-T}K A^{-1}$ matrices in the expression above.
To do so it is useful to define the following matrix
\begin{equation}
 \widetilde T=K^{1/2}T K^{1/2},
\end{equation}
which is symmetric and positive definite and therefore can be diagonalized by an orthogonal matrix $\mathcal{O}$. 
Without loss of generality, this orthogonal matrix $\mathcal{O}$ can be parametrized as
\begin{equation}
\label{omatrix}
\mathcal{O}=\left(\begin{matrix}\cos\theta &-\sin \theta\\\sin \theta &\cos \theta\end{matrix}\right),
\end{equation}
and choosing the parameter $\theta$ (not to be confused with the angles  $\theta_i$) by
\begin{equation}
\label{tan2theta}
 \tan2\theta=\frac{2 l_1}{(l_1-l_2)}\sqrt{\frac{m_2l_2}{Ml_1}},
\end{equation}
we have that
\begin{equation}
 \mathcal{O}^T \widetilde T \mathcal{O}=\text{diag}(\omega_1^2,\omega_2^2)=T_d.
\end{equation}
The eigenvalues $\omega_i^2$ are positive since $\widetilde T$ is a positive definite matrix,  and have 
dimensions of (angular) frequency square. The explicit expressions are
\small
$$
 \omega_1^2=\frac{g}{m_1l_1l_2}\left(-\sqrt{Mm_2l_1 l_2}\sin 2\theta +Ml_1 \sin ^2\theta+Ml_2 \cos ^2\theta\right),
$$
$$
\hspace{-.22cm}\omega_2^2=\frac{g}{m_1l_1l_2} \left(\sqrt{Mm_2l_1 l_2}\sin 2\theta +Ml_1 \cos ^2\theta+Ml_2 \sin ^2\theta\right),
$$
\normalsize
in agreement with the eigenfrequencies given  in \cite{Blevins1979}.
Now, by writing the transformation matrix as 
\begin{eqnarray}
\label{Amatrix}
\hspace*{-.4cm} A&=&\mathcal{O}^TK^{1/2}=
 \left(\begin{matrix}
  \sqrt{M g l_1} \cos \theta &  \sqrt{m_2 g l_2} \sin \theta \\
 -\sqrt{M g l_1}\sin \theta & \sqrt{m_2 g l_2} \cos \theta
\end{matrix}\right)\!,
\end{eqnarray}
%
both quadratic terms in the transformed Hamiltonian (\ref{Hu_coupled}) are diagonal since
\begin{eqnarray}
  ATA^T&=&\mathcal{O}^TK^{1/2} T K^{1/2}\mathcal{O}=\mathcal{O}^T\widetilde T \mathcal{O}=T_d,\\
  A^{-T}KA^{-1}&=&\mathcal{O}^TK^{-1/2}KK^{-1/2}\mathcal{O}=1.
\end{eqnarray} 
%
%
%
%

Finally, 
the Hamiltonian (\ref{Hu_coupled}) takes the uncoupled form
\begin{eqnarray}
 H_u&=&\frac{1}{2}\sum_{i=1}^2 \left(\omega_i^2 p_{u_i}^2+u_i^2\right)
 \nonumber\\
 &+&\dot x\sqrt{\frac{Mg}{l_1}}(-p_{u_1} \cos\theta+p_{u_2}\sin\theta).
 \label{H_u_uncoupled}
\end{eqnarray}
%

\subsection{Lewis-Leach family of Hamiltonians and second canonical transformation}
The  Lewis-Leach (LL) family  
  of Hamiltonians are of the form \cite{LewisLeach1982} 
   \begin{equation}
   \label{LLfamily}
    H_{LL}= \frac{1}{2}\left[p^2 +\Omega(t) q^2\right] - F(t) q,
   \end{equation}
  i. e.,  quadratic Hamiltonians with linear in position terms. For them quadratic invariants are explicitly known. 
By a suitable canonical transformation to some  generalized coordinates $\{ q_i,p_i\}$, we shall transform  $H_u$ into 
this form.  
   This can be easily achieved just by exchanging momentum and coordinate  \cite{Dittrich2001}. 
   The transformation is generated by   $F_1=u_1 q_1+u_2 q_2$  which 
   gives the new coordinates and momenta in terms of the  old ones as follows, 
   \begin{eqnarray}
  q_i  &=&\frac{\partial F_1}{\partial u_i}=p_{u_i},\\
    p_{i}&=&-\frac{\partial F_1}{\partial q_i}=-u_i.
   \end{eqnarray}
By using this canonical transformation, the new Hamiltonian is 
\begin{eqnarray}
 H_q&=&H_1+H_2,
 \label{Hq}
\end{eqnarray}
a sum of two independent forced harmonic oscillators that  belong to the LL family,
\begin{eqnarray}
 H_1&=&\frac{1}{2} \left(p_1^2+\omega_1^2 q_1^2\right) -q_1 \dot x\sqrt{\frac{Mg}{l_1}} \cos\theta,
 \nonumber \\
 H_2&=&\frac{1}{2} \left(p_2^2+\omega_2^2 q_2^2\right) +q_2\dot x\sqrt{\frac{Mg}{l_1}}\sin\theta.
 \label{H1&H2}
\end{eqnarray}
%
%
%
%
 \subsection{Explicit expression of normal mode coordinates}
 Taking into account the two canonical transformations, the explicit expression of the normal mode coordinates and momenta $\{q_i,p_i\}$ in terms of  the original 
variables $\{\theta_i,p_{\theta_i}\}$ is
 \begin{eqnarray}
  \left(\begin{matrix}q_1\\q_2\\p_1\\p_2\end{matrix}\right)&=&
  \left(\begin{matrix}0&{\cal I}_2\\-{\cal I}_2&0\end{matrix}\right)
  \left(\begin{matrix}A&0\\0&A^{-T}\end{matrix}\right)
  \left(\begin{matrix}\theta_1\\\theta_2\\p_{\theta_1}\\p_{\theta_2}\end{matrix}\right),
 \end{eqnarray}
where ${\cal I}_2$ is the $2\times 2$ identity matrix  
and, using the explicit expression of $A$  in (\ref{Amatrix}), we have 
 \begin{eqnarray}
  q_1&=& \frac{\cos \theta}{\sqrt{Mgl_1}} p_{\theta_1}+ \frac{\sin \theta}{\sqrt{m_2 g l_2}} p_{\theta_2},
  \nonumber\\
  q_2&=& - \frac{\sin \theta}{\sqrt{Mgl_1}}p_{\theta_1}+\frac{\cos \theta}{\sqrt{m_2 g l_2}} p_{\theta_2},
\nonumber\\
p_1&=&-\cos \theta\sqrt{Mg l_1} \theta_1-\sin \theta \sqrt{m_2g l_2}  \theta_2,
\nonumber\\
 p_2&=&  \sin \theta\sqrt{Mg l_1}\theta_1- \cos \theta \sqrt{m_2g l_2}\theta_2.
 \end{eqnarray}

 \section{Designing the STA protocol}
 \label{sec_sta}
We are now ready to define the invariants and design the 
driving function $x(t)$. 
%
 %
\subsection{Dynamical invariants}
A dynamical invariant of a Hamiltonian system remains constant during the time evolution  \cite{Damour2000}.
Labelling the dynamical invariant of the  Hamiltonian $H_i$ as $I_i$ we have that
\begin{eqnarray}
 \frac{dI_i}{dt}&=&\partial_t I_i+\{I_i,H_i\}=0,
\end{eqnarray}
with $\{I_i,H_i\}$ being the Poisson bracket.
The sum of invariants $I=I_1+I_2$ is invariant with respect to  the sum of Hamiltonians $H_q=H_1+H_2$ since
 \begin{eqnarray}
  \frac{dI}{dt}&=&\{I,H_q\}+\partial_tI=\{I_1+I_2,H_1+H_2\}+\partial_t(I_1+I_2)\nonumber\\
  &=&\left(\{I_1,H_1\}+\partial_tI_1\right)+\left(\{I_2,H_2\}+\partial_tI_2\right)\nonumber\\
  &+&\{I_1,H_2\}+\{I_2,H_1\}=0\nonumber
 \end{eqnarray}
%
The invariants for (\ref{H1&H2})  have the explicit form \cite{LewisLeach1982}
 \begin{eqnarray}
  I_i&=&\frac{1}{2}\left(p_i-\dot \alpha_i\right)^2+\frac{\omega_i^2}{2}\left(q_i-\alpha_i\right)^2,
\label{invar}
 \end{eqnarray}
 provided the functions $\alpha_i$ satisfy the following Newton equations,
 \begin{eqnarray}
 \label{newton_eq_1}
  \ddot \alpha_1+\omega_1^2\alpha_1&=&\dot x\sqrt{\frac{Mg}{l_1}} \cos\theta,\\
  \ddot \alpha_2+\omega_2^2\alpha_2&=&-\dot x\sqrt{\frac{Mg}{l_1}}\sin\theta.
  \label{newton_eq_2}
 \end{eqnarray}
These $\alpha_i$ functions may be regarded as auxiliary, reference, special  ``displacements'' in two forced harmonic oscillators. 
Let us underline that the actual motion 
for a specific transport process is described by the $q_i$ rather than by the $\alpha_i$
(which represent just a particular case of all possible $q_i$). 
Note by the way that the $q_i$ satisfy the same Newton equations (with the same forces) as the $\alpha_i$. 
However, we shall impose to $\alpha_i$ boundary conditions that will guarantee zero final excitations 
whereas the initial conditions for the $q_i$ are arbitrary.      

%
\subsection{Boundary conditions (BC) for $x(t)$ and $\alpha_i(t)$}
We shall assume a transport from $x(0)=0$ to $x(t_f)=d$ with additional  smooth 
boundary conditions for the trolley velocity,   $\dot x(t_b)=0$ for $t_b=0,t_f$. 
We shall further assume that the auxiliary functions $\alpha_i$, as well as their first and second time derivatives vanish at boundary times $t_b=0,t_f$. 
We therefore have in principle a total of sixteen boundary conditions (BC), namely
\begin{eqnarray}
  \alpha_i(t_b)&=&\dot \alpha_i(t_b)=\ddot \alpha_i(t_b)=0,
  \nonumber\\
  x(0)&=&0\textrm{ ; }  x(t_f)=d,
  \nonumber\\
  \dot x(0)&=&0\textrm{ ; }\dot x(t_f)=0.
  \label{BC_eqs}
 \end{eqnarray}
%
%
These boundary conditions guarantee that each invariant $I_i$ coincides with the corresponding
Hamiltonian $H_i$ at initial and final times,  see  (\ref{invar}), 
\begin{eqnarray}
 H_q(t_b)=H_1(t_b)+H_2(t_b)=I_1(t_b)+I_2(t_b)=I(t_b).\nonumber 
\end{eqnarray}
At these boundary times, and due to the $\dot x(t_b)=0$ boundary condition,  the  Hamiltonian 
represents the total mechanical energy of the system, i. e. $E(t_b)=H_q(t_b)$. 
If a fast finite-time process is designed so that the auxiliary functions $\alpha_i$ satisfy the imposed boundary conditions, 
the energy at final and initial times -regardless of the initial conditions of the hook and load, i.e., regardless of the 
initial conditions set for $q_i(0)$ and its derivatives- will coincide since
$$
E(0)=H_q(0)=I(0)=I(t_f)=H_q(t_f)=E(t_f).
$$
Note that in principle the only conditions needed to guarantee $I(t_b)=H(t_b)$ are the ones for $\alpha(t_b)$ and 
$\dot{\alpha}(t_b)$. The others have a physical motivation as the desired boundaries for the trolley motion (on $x(t_b)$ and $\dot{x}(t_b)$) or are a consequence of the former ones (the ones on $\ddot{\alpha}(t_b)$ because of (\ref{newton_eq_1}) and (\ref{newton_eq_2})).  

In the following  subsection we will show how to construct the trolley trajectory $x(t)$ so that the desired conditions 
in  (\ref{BC_eqs}) are satisfied.
%
%
%

\subsection{Inverse engineering}
\label{inverse_eng_sec}
%
%
%
\begin{figure}[t]
\begin{center}
\includegraphics[width=8.5cm]{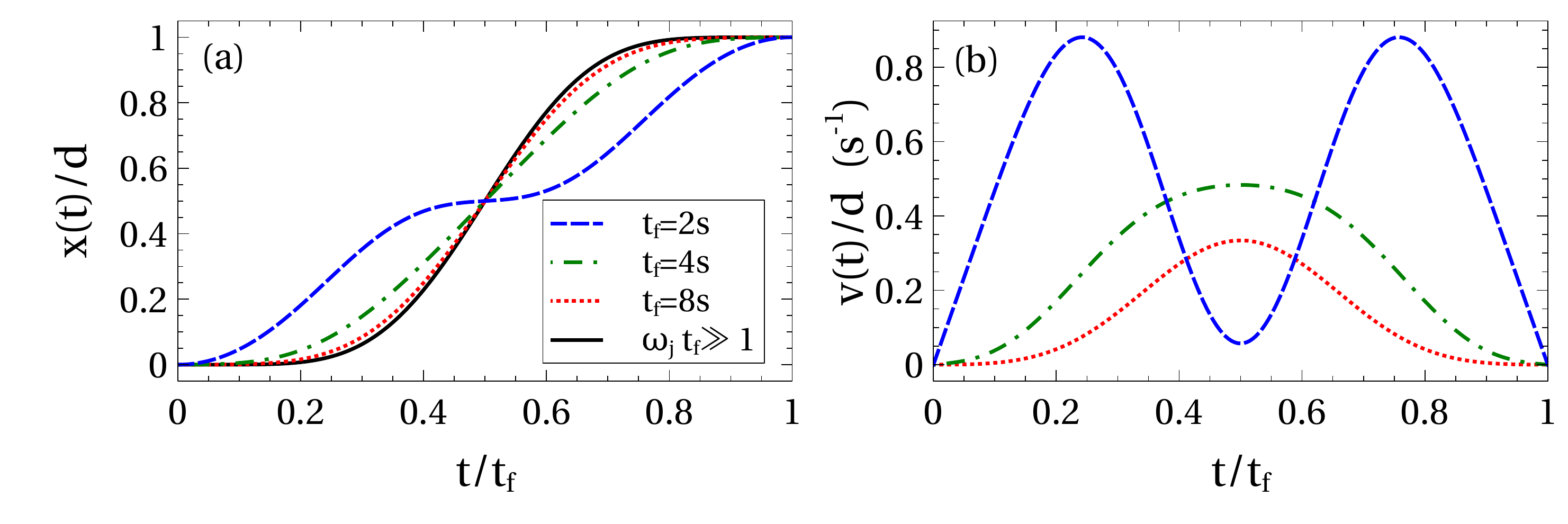}
\caption{(Color online) Trolley trajectories $x(t)$  and velocities $v(t)=\dot x(t)$ for different final times: 
$t_f=2s$ (blue-dashed line), $t_f=4s$ (green-dot-dashed line), $t_f=8s$ (red-dotted line). Compare to the ``long time behaviour'' 
in (a) of (\ref{xt_inf}) (black-solid line).
Other parameters are: $m_1=1$ kg, $m_2=0.5$ kg, $l_1=1$ m, $l_2=0.2$ m.}
\label{fig_trolley}
\end{center}
\end{figure}
%
We start by proposing the following  ansatz for the trolley velocity $\dot x(t)$, symmetric with respect to $t_f/2$, 
\begin{equation}
\label{xdot_ansatz}
\dot x(t)=\sum_{j=1}^{3} a_j \sin{\frac{(2j-1)\pi t}{t_f}},
\end{equation}
with three free parameters $a_1$, $a_2$,  and $a_3$  that will be determined from the following three conditions (the second line involves two conditions, 
one for each frequency, as justified in the Appendix):
\begin{eqnarray}
\int_0^{t_f} \dot x(\tau)d\tau&=&d,\\
\int_0^{t_f} \dot x (\tau) \cos[\omega_j(\tau-\frac{t_f}{2})] d\tau&=&0,
\label{3conditions}
\end{eqnarray}
for $j=1,2$. 
\begin{figure}[t!]
\begin{center}
\includegraphics[width=8.5cm]{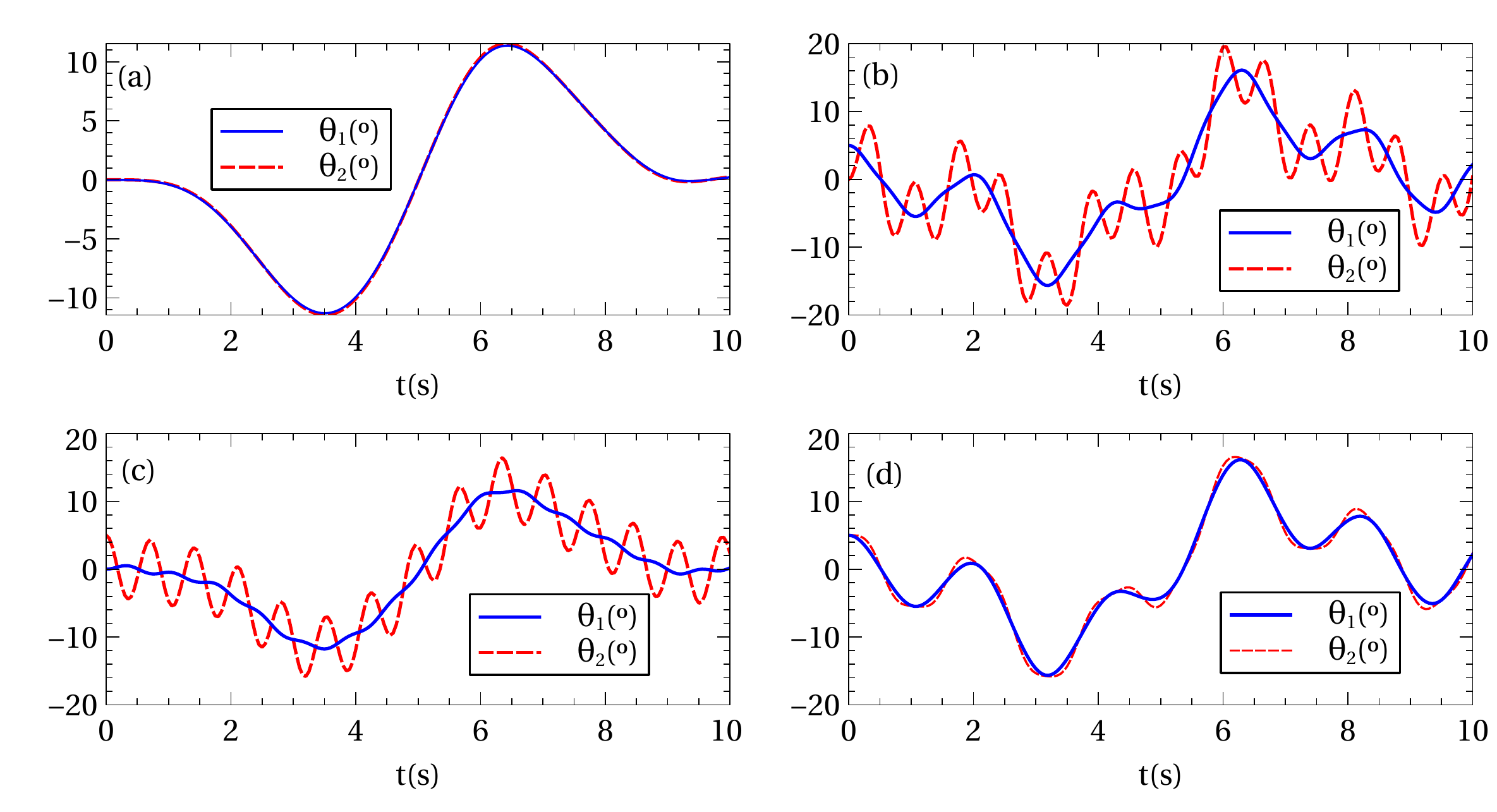}
\caption{(Color online) Time evolution of the suspension angles for a transport of $d=15$ m in a time $t_f=10$ s.
We have numerically integrated the exact dynamical equations using the exact Lagrangian (\ref{lagrangian_exact}) with different initial conditions:
(a) $\theta_1(0)=0º$, $\theta_2(0)=0º$; 
(b) $\theta_1(0)=5º$, $\theta_2(0)=0º$; 
(c) $\theta_1(0)=0º$, $\theta_2(0)=5º$; and
(d) $\theta_1(0)=5º$, $\theta_2(0)=5º$,
with $\dot \theta_1(0)=\dot \theta_2(0)=0$ in all cases.
In the scale of the figure the results using  the approximate Lagrangian (\ref{lagrangian_approx}) 
or the exact one are indistinguishable. 
Other parameters are: $m_1=1$ kg, $m_2=0.5$ kg, $l_1=1$ m, $l_2=0.2$ m.}
\label{fig_angles}
\end{center}
\end{figure}
Different functional forms are possible, but this ansatz is chosen for simplicity and because of very useful properties discussed 
in the Appendix 
(it avoids resonance and boundary effects).  
It is also remarkable that an ansatz with only three free parameters satisfies the full set of sixteen boundary conditions  in (\ref{BC_eqs}),
see further details in the Appendix. 

The three free parameters can be therefore written in terms of 
the system physical parameters as
\begin{eqnarray}
 a_1&=&\frac{75 \pi  d \left( \omega_1^2 t_f^2-\pi ^2\right) \left( \omega_2^2 t_f^2-\pi ^2\right)}{128 t_f^5 \omega_1^2 \omega_2^2},\\
 a_2&=&-\frac{75 \pi  d \left( \omega_1^2 t_f^2-9 \pi ^2\right) \left( \omega_2^2 t_f^2-9 \pi ^2\right)}{256 t_f^5 \omega_1^2 \omega_2^2},\\
 a_3&=& \frac{15 \pi  d \left( \omega_1^2 t_f^2-25 \pi ^2\right) \left( \omega_2^2 t_f^2-25 \pi ^2\right)}{256 t_f^5 \omega_1^2 \omega_2^2}.
 \label{aj_parameters}
\end{eqnarray}
These parameters determine completely the velocity of the trolley  by (\ref{xdot_ansatz}), and its trajectory is simply the integral 
\begin{eqnarray}
\label{trolley_traj}
 x(t)=\int_0^t \dot x(\tau)d\tau,
\end{eqnarray}
which gives  an explicit but lengthy expression. See some trolley trajectories and velocities in Fig. \ref{fig_trolley}. 
For long transport times ($\omega_jt_f\gg \pi$) the trolley trajectory becomes independent of the masses and lengths of the pendulum and tends to
\begin{eqnarray}
\label{xt_inf}
 x_\infty(t)&=&d\left[\frac{1}{2}-\frac{75}{128} \cos \left(\frac{\pi  t}{t_f}\right)+\frac{25}{256} \cos \left(\frac{3 \pi    t}{t_f}\right)\right.\nonumber\\
 &-&\left.\frac{3}{256} \cos \left(\frac{5 \pi  t}{t_f}\right)\right].
\end{eqnarray}
This trajectory implies a  maximal velocity  $v_{max}=({15\pi}/{16})({d}/{t_f})$ at $t=t_f/2$.
In this asymptotic scenario  there is only one acceleration time segment up to $t_f/2$ and a subsequent braking segment.

For short times compared to eigenperiods there are several segments of acceleration and braking. 
 In any case this regime is less interesting in practice since the system deviates from the harmonic regime.

%
%
\section{Numerical results}
\label{sec_results}
%
\subsection{Time evolution of suspension angles}

Once the trolley trajectory is designed, the dynamical evolution of the system can be found  by numerically integrating
the Euler-Lagrange equations of motion using either the exact Lagrangian (\ref{lagrangian_exact}) or the approximate Lagrangian in the harmonic (small oscillations) approximation (\ref{lagrangian_approx}).
In Fig. \ref{fig_angles} some examples of the time evolution of the suspension angles $\theta_1$ and $\theta_2$ during transport are shown.
The initial and final angles are not equal (unless the system is initially at equilibrium), but this is not a requirement for ending with the initial energy. 
%
The calculation has been done using the exact Lagrangian, but the results are undistinguishable in the scale of the figure when using the  approximate
Lagrangian since the involved angles are small throughout the whole transport process.
For larger transport distance $d$ or smaller process time $t_f$ these differences will increase  and will lead to some errors due to the anharmonicity of the exact model
as will be discussed in the following section.

%
%

%
\subsection{Anharmonic effects}
%
\begin{figure}[t!]
\begin{center}
\includegraphics[width=7.5cm]{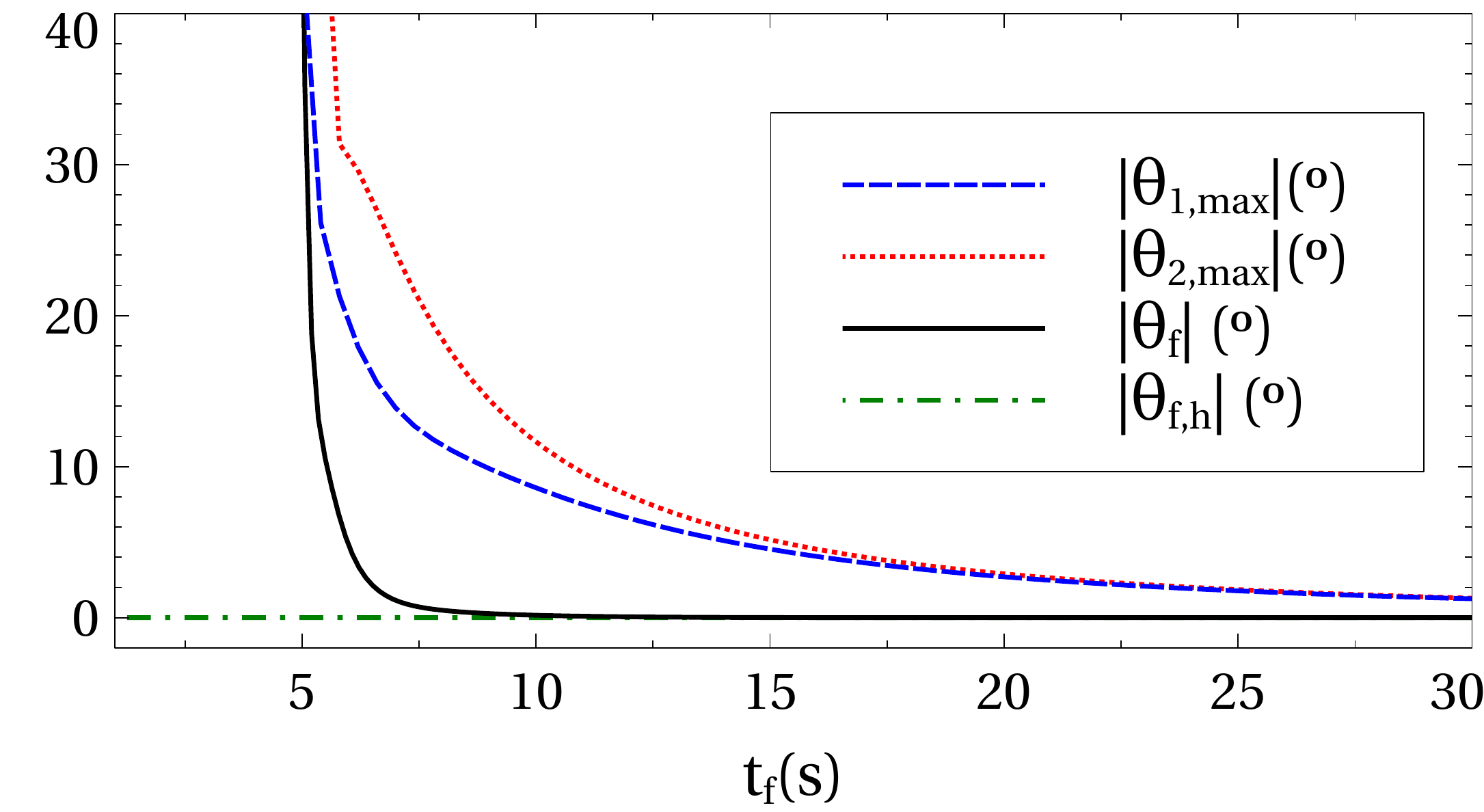}
\caption{(Color online)
Maximum swing angles during the process as a function of the duration $t_f$ (red dotted and blue dashed lines).
For very rapid operations (small $t_f$), larger angles are involved and the harmonic approximation breaks down.
Fictitious angle $\theta_f$ (black-solid line), which basically is a measure of the final excitation energy, see the main text,  as a function of $t_f$.
In the harmonic approximation this angle is zero by construction ($\theta_{f,h}$, green-dashed-dotted line). 
System assumed initially in equilibrium.
Other parameters are: $m_1=30$ kg, $m_2=3$ kg, $l_1=30$ m, $l_2=3$ m, $d=15$m.
The natural periods of the modes are $T_1=2\pi/\omega_1=11.048$ s and $T_2=2\pi/\omega_2=3.298$ s.}
\label{fig_error_tf}
\end{center}
\end{figure}

For rapid transport operations, 
the involved angles are larger and the harmonic approximation breaks down, see Fig. \ref{fig_error_tf}. 
Therefore, some deviations from the ideal results (i. e., equal final and initial energies) should be expected.

To quantify the excitation at final time in a way that is easy to understand and visualize,
we measure the final energy $\Delta E$ in terms 
of a fictitious angle $\theta_f$. This angle is defined as follows: 
{\it (i)} the load and hook are initially in equilibrium (at rest in the vertical position); and 
{\it (ii)} the final energy is artificially interpreted as pure potential energy 
for a configuration where load and hook are at rest along a line with  $\theta_f=\theta_1=\theta_2$. 
In other words: $\theta_f$ is the final angle when the final energy is considered to be purely potential and the two 
suspension angles coincide. Using (\ref{exact_TV}) we may write
\begin{eqnarray}
 \Delta E=-2E_0 \sin ^2\frac{\theta_f }{2}.
 \label{theta_f}
\end{eqnarray}
with $E_0=- m_1 g l_1- m_2 g (l_1+l_2)$ being the energy for the  equilibrium configuration.
In Fig. \ref{fig_error_tf}, this fictitious angle is plotted as a function of the process duration time $t_f$.

%
\subsection{Stability}

\begin{figure}[t!]
\begin{center}
\includegraphics[width=8.5cm]{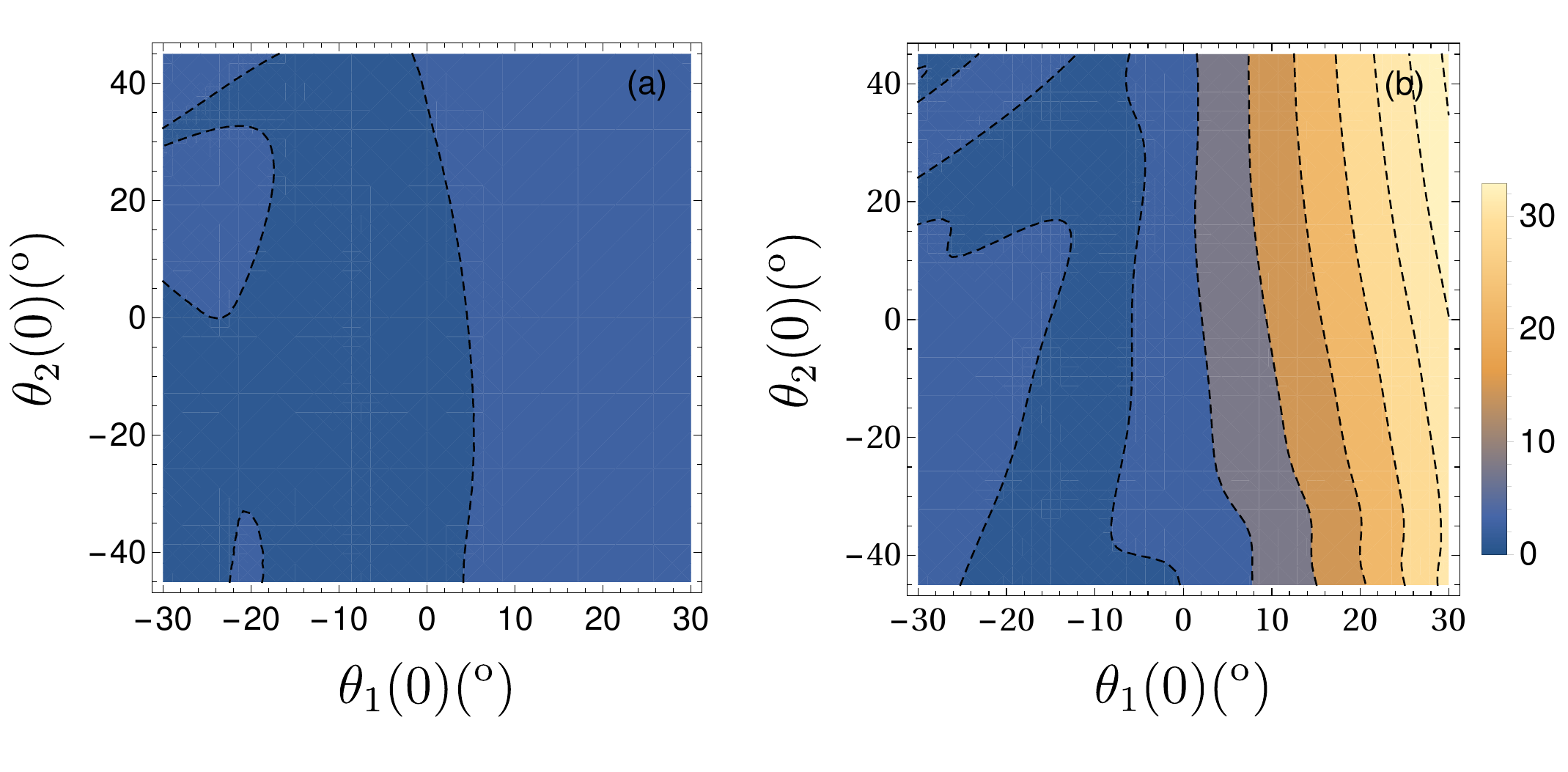}
\caption{(Color online) Difference between final and initial energy measured by the modulus of the fictitious angle $\theta_f$ (in º)
 as a function of the deviations from equilibrium configuration of either $\theta_1(0)$ or $\theta_2(0)$
after solving the exact dynamics with Lagrangian (\ref{exact_TV}). 
(a) Final fictitious angle for the inverse-engineered trolley trajectory  (\ref{trolley_traj}).
(b) Final fictitious angle for the  postulated cubic trajectory   (\ref{3rdorder_xt}).
Other parameters are: $m_1=1$ kg, $m_2=0.5$ kg, $l_1=1$ m, $l_2=0.2$ m, $d=15$ m and $t_f=5$ s.
The system is assummed initially at rest, $\dot \theta_1(0)=\dot \theta_2(0)=0$.}
\label{fig_stability_contour}
\end{center}
\end{figure}
%
The stability of the proposed transport protocol can be studied by allowing some initial deviations 
of the angles $\theta_1(0)$ or $\theta_2(0)$ from the equilibrium positions. 
In Fig. \ref{fig_stability_contour}a, the final time energy excitation, measured in units of the fictitious angle $\theta_f$ (\ref{theta_f}), 
is plotted as a function of these deviations.

We will compare the resulting excitation with that for a simple third order polynomial ansatz for the trolley trajectory,
\begin{eqnarray}
\label{3rdorder_xt}
 x(t)=3 d\left(\frac{t}{t_f}\right)^2-2 d\left(\frac{t}{t_f}\right)^3,
\end{eqnarray}
which  satisfies the four BCs in (\ref{BC_eqs}) for $x(t)$ but not those for the auxiliar functions $\alpha_i$.
As shown in Fig. \ref{fig_stability_contour}b (which should be compared with Fig. \ref{fig_stability_contour}a),
the excitation at final time using this simple trajectory  is much larger that 
the one using the inverse engineered trajectory.
Our inverse engineering method leads to much more robust results.
%

\subsection{Example limiting the maximal trolley speed}
\label{sec_actual_crane}
The engine power and safety considerations imply limits to the trolley speed. 
In this example we test the effect of such a limit. 
We set a load  $m_2=1000$ kg  transported 
a distance $d=40$ m. We also set a hook mass  $m_{1}=150$ kg,   $l_{2}=5$ m, and  
$l_{1}=40$ m. A  maximum velocity of $2$ m/s is assumed.

\begin{figure}[t!]
\begin{center}
\includegraphics[width=8.5cm]{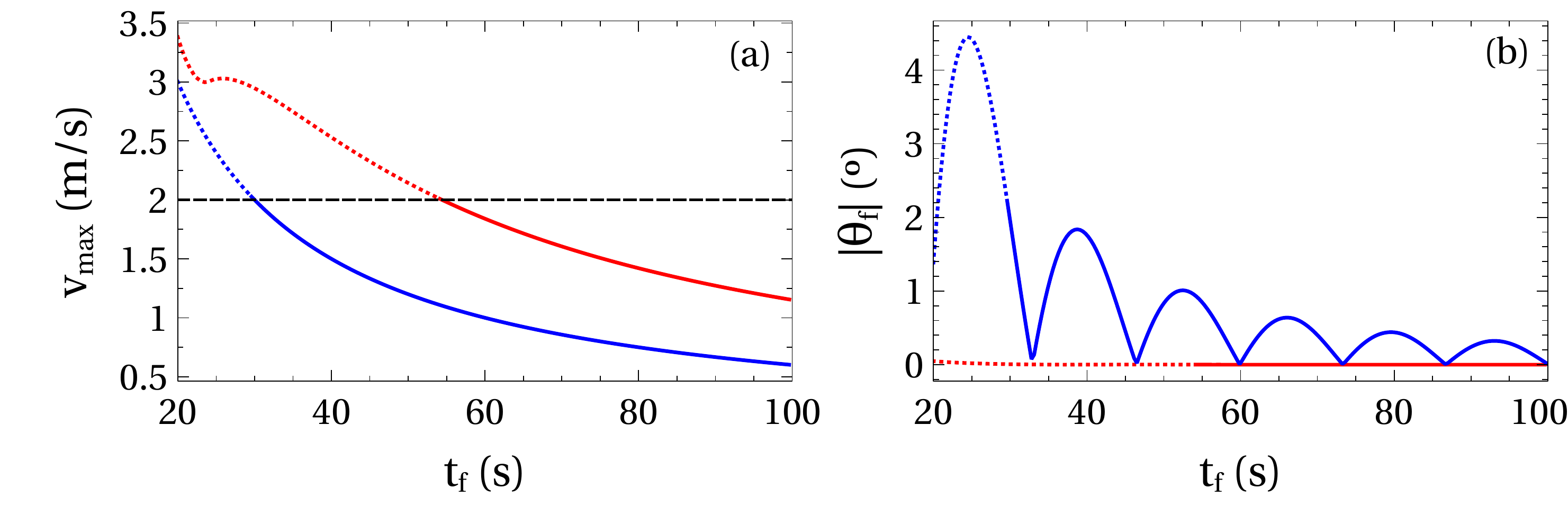}
\caption{(Color online) Comparison of two transport protocols,
the inverse engineered trolley trajectory  (\ref{trolley_traj}) (red) and 
the cubic trajectory   (\ref{3rdorder_xt}) (blue).
The dotted line is for segments where the maximal trolley velocity is larger than $2$ m/s, whereas in solid line
segments the maximal velocity is below that value. 
(a) Maximum trolley velocity during the process and (b) excitation at final time measured by the fictitious angle $\theta_f$.
Rest of parameters: $m_1=150$ kg, $m_2=1000$ kg, $l_1=40$ m, $l_2=5$ m and $d=40$ m. System initially at equilibrium.
The natural periods of the modes are $T_1=2\pi/\omega_1=13.376$ s and $T_2=2\pi/\omega_2=1.538$ s.}
\label{fig_liehberr}
\end{center}
\end{figure}

With this data,  two transport  protocols are compared in Fig. \ref{fig_liehberr}:
(i) inverse engineered trolley trajectory (\ref{trolley_traj}) and
(ii) directly postulated cubic trajectory  (\ref{3rdorder_xt}).
For initial conditions at equilibrium and the same final process time $t_f$, our inverse engineering protocol involves higher
maximum velocities but the crane ends with much lower energy, almost ending in equilibrium.
In the dotted part of the curves the limit of $2$ m/s
is surpassed.

\section{Conclusions}
\label{sec_conclussions}

We have applied an invariant based inverse engineering  STA method to design fast trolley trajectories of a double pendulum overhead crane. In the small oscillations regime 
these trajectories guarantee that the transport does not induce any energy excitation, regardless of the initial condition of the double pendulum. 
We have first found the normal modes and from them the dynamical invariants. 
Using these invariants, it is possible to inverse engineer STA trolley trajectories. 
We have performed the numerical simulations with the exact dynamics  to see the  parameter intervals  where the protocol is accurate. 
Comparisons are also made with less sophisticated trolley trajectories that demonstrate the advantage of the 
STA approach.  
We have worked out a particularly simple design for the trolley speed with three sine terms, (\ref{xdot_ansatz}). It should be clear that we have not really optimized the trolley trajectory. One of the interesting facts about STA methods is that the solutions to the inverse problem are not unique. That means that there is much room for finding specific trajectories that 
optimize  variables of interest, or are robust with respect to specific perturbations or parametric uncertainties \cite{Ruschhaupt2012}. STA combine well in particular with optimal control theory \cite{review2019}. Thus STA provide a useful avenue to minimize the sensitivity to parameter uncertainties, one of the weak 
points of open-loop approaches.  
Other possible extension of this work may be to tackle combined or sequential operations with transport and hoisting \cite{Gonzalez-Resines2017}.  
 
Compared to previous work on methods without feedback \cite{Masoud2014,Vaughan2010,Hong2003}, this paper exemplifies and introduces the use of shortcuts-to-adiabaticity 
in mechatronics for multimode systems.  We refrain from performing a numerical comparison with ``input shaping'' methods because virtually any result would be  possible given the flexibility of both input-shaping and STA  methods to accomodate  a vast family of possible designs for the trolley motion, corrections for increased robustness with respect to parameter uncertainties or  noise. Nevertheless we would like to underline the simplicity 
of the basic invariant-based engineering for the  
moving double pendulum crane, compared to input-shaping approaches \cite{Masoud2014,Vaughan2010,Hong2003}. 
Even if the choice among methods may be a matter of taste 
and previous experience, we would like to argue that STA should be in the the toolbox 
of control methods, if only because  STA are well tested and have been intensely developed theoretically 
along different approaches and applied to many experiments in AMO (atomic, molecular, and optical), and solid state 
physics \cite{review2019}. Thus engineering applications may benefit from an important 
framework of techniques and concepts.  By the way, a positive influence in the opposite direction, from mechatronics to AMO physics, is also expected. For example, state manipulation  in AMO science has much to learn from a long experience on  control with feedback in mechatronics.

\acknowledgments{We thank B. Ruiz L\'opez and S. Gonz\'alez-Resines for collaboration during their final-degree and master Theses.  
This work was supported by the Basque Country Government (Grant No. IT986-16) and PGC2018-101355-B-I00 (MCIU/AEI/FEDER,UE).}

\appendix
\section{Simple ansatz for trolley velocity}
\label{constraints_app}
The ansatz with three free parameters (\ref{xdot_ansatz}),
\begin{equation}
\label{a1}
 \dot x(t)=\sum_{k=1}^{3} a_k \sin{\frac{(2k-1)\pi t}{t_f}},
\end{equation}
is an even function with respect to $t=t_f/2$ and  it automatically satisfies  $\dot x(t_b)=0$. 
We now rewrite  the auxiliary equations (\ref{newton_eq_1}) and (\ref{newton_eq_2}) as
\begin{eqnarray}
\label{beta_eqs}
\ddot \beta_j+\omega_j^2\beta_j&=&\dot x
\end{eqnarray}
for $j=1,2$,  $\beta_1=\alpha_1 \sqrt{{Mg}/{l_1}} \cos\theta$, and $\beta_2=-\alpha_2 \sqrt{{Mg}/{l_1}}\sin\theta$.
These are the equations of a driven harmonic oscillator, where the trolley velocity plays the role of the external driving force.
The new variables $\beta_j$ should satisfy the BC  $\beta_j(t_b)=\dot \beta_j(t_b)=\ddot \beta_j(t_b)=0 $, see (\ref{BC_eqs}).

The solutions with initial condition  $\beta_j(0)=\dot\beta_j(0)=0$ to the above equations (\ref{beta_eqs}) 
can be written in a compact form using the complex function \cite{interfermoetry2018}
\begin{eqnarray}
\label{beta_sols}
 z_j(t)&=&\omega_j \beta_j(t)+i \dot\beta_j(t)\nonumber\\
 &=&ie^{-i\omega_jt}\int_0^t \dot x(\tau) e^{i\omega_j \tau}d\tau.
\end{eqnarray}
They also satisfy $\ddot\beta_j(0)=0$ because  $\dot x (0)=0$, see  (\ref{beta_eqs}).
If we now impose that the integral term in the expression above vanishes at $t=t_f$ for $j=1,2$,
the final time BCs  $\beta_i(t_f)=\dot \beta_i(t_f)=\ddot \beta_i(t_f)=0 $ will be also satisfied. 
Note that since $\dot x(\tau)$ is an even function in the integration interval, we may rewrite 
the integral above as
%
\begin{eqnarray}
 \int_0^{t_f} \dot x(\tau) e^{i\omega_j\tau}d\tau=e^{i\omega_j t_f/2}\int_0^{t_f} \dot x (\tau) \cos\left[\omega_j\left(\tau-\frac{t_f}{2}\right)\right] d\tau\nonumber
\end{eqnarray}
so that only the cosine part of the remaining exponential may contribute by symmetry.

We have therefore two integral constraints, but three parameters to determine in (\ref{a1}).
The third, and last, constraint comes from the fact that the crane trajectory ends at $x(t_f)=d$,
\begin{eqnarray}
x(t_f)= \int_0^{t_f} \dot x(\tau)d\tau&=&d.
\end{eqnarray}
The three conditions are therefore those summarized in  (\ref{3conditions}).

In the steps just described the sixteen boundary conditions in (\ref{BC_eqs}) are satisfied. 
Let us count them explicitly: 
two for $\dot{x}(t_b)$, four for the initial conditions set for $\beta_j(0)$ and $\dot{\beta}_j(0)$;  
two more because $\ddot{\beta}(0)$ vanishes automatically due to $\dot{x}(0)=0$;  
four are satisfied for the  $\beta_j$ and their first derivatives at $t_f$ when nullifying the integral; this implies two more, 
$\ddot{\beta_j}(t_f)=0$ using $\dot{x}(t_f)=0$; finally 
the last two correspond to the initial condition $x(0)=0$  and final condition $x(t_f)=0$.

Equation (\ref{beta_sols}) is also useful to analyze possible unstable behavior due to border effects.
For the proposed ansatz for the trolley velocity (\ref{a1}) 
the only discontinuity will be  in the initial (and final) acceleration but it behaves nicely with $t_f$, 
\begin{eqnarray}
 \ddot x(0)&=&-\ddot x(t_f)=\frac{225d\pi^6}{2\omega_1^2\omega_2^2t_f^{6}}.
\end{eqnarray}
(The effect of $\ddot x(0)$ on the $\beta_j$ as a boundary term
is made evident by integrating (\ref{beta_sols})
twice by parts.) 
We tried other ansatzes, polynomials in particular, with a much worse behavior and serious boundary-driven  excitations because of periodic singularities of $\ddot{x}$ at the boundary times. 

If a null acceleration is imposed at boundary times, a further term in (\ref{xdot_ansatz})
will be needed and a 
discontinuity in the fourth derivative of $x(t)$ scaling as $t_f^{-8}$ will be observed. Every odd derivative of a series like (\ref{a1}) is zero by construction, regardless of the number of terms.
In general, imposing a null $2n$th derivative of $x(t)$ leads to a discontinuity in the $(2n+2)$th derivative, scaling as $t_f^{-(2n+6)}$, i. e., 
\begin{eqnarray}
 x^{(2n)}(t_b)=0\rightarrow  x^{(2n+2)}(t_b)\propto t_f^{-(2n+6)}. 
\end{eqnarray}

As for the potential occurrence of resonances because of matching of frequencies in 
(\ref{beta_sols}), our ansatz (\ref{a1}) leads to very stable and simple $\beta_j$ forms without any  resonant behavior,  in particular no dangerous denominators arise.   
Thus our trolley trajectory avoids resonances from both 
modes in a simple direct way, for other methods using multi-mode input shaping see \cite{Masoud2014,Vaughan2010}.

\bibliography{mybib.bib}{}
%

%

\end{document}